\documentclass[prl,twocolumn,showpacs,preprintnumbers]{revtex4}
\usepackage{amssymb}

\usepackage{graphicx}
\usepackage{dcolumn}
\usepackage{bm}

\begin{document}

\title{Ferromagnetism and Metal-Insulator Transition in the Disordered
Hubbard Model }
\author{Krzysztof Byczuk,$^{1,2}$ Martin Ulmke,$^{3,1}$ and Dieter Vollhardt$%
^1$ }
\affiliation{\centerline{$^1$Theoretical Physics III, Center for Electronic Correlations and
Magnetism, Institute for Physics,} \centerline{University of Augsburg,
D-86135 Augsburg, Germany }
\centerline {$^2$ Institute of Theoretical Physics,
Warsaw University, ul. Ho\.za 69, PL-00-681 Warszawa, Poland}
\centerline{$^3$
  FGAN - FKIE,
   Neuenahrer Stra\ss{}e 20   
  D-53343 Wachtberg, Germany} }
\date{\today}

\begin{abstract}
A detailed study of the paramagnetic to ferromagnetic phase
transition in the one-band Hubbard model in the presence of binary
alloy disorder is presented. The influence of the disorder (with
concentration $x$ and $1-x$ of the two alloy ions) on the Curie
temperature $T_{c}$ is found to depend strongly on electron
density $n$. While at high densities, $n>x$, the disorder always
reduces $T_c$, at low densities, $n<x$, the disorder can even
\emph{enhance} $T_c$ if the interaction is strong enough. At the
particular density $n=x$ (i.\,e.\ not necessarily at half filling) the 
interplay between disorder-induced band splitting and correlation
induced Mott transition gives rise to a new type of
metal-insulator transition.
\end{abstract}

\pacs{71.10.-w,71.10.Fd,71.27.+a,75.40.Cx  }
\maketitle





In correlated electron materials it is a rule rather than an exception that
the electrons, apart from strong interactions,  are also subject to
disorder. The disorder may result from non-stoichiometric composition, as
obtained, for example, by doping of manganites (La$_{1-x}$Sr$_{x}$MnO$_{3}$)
and cuprates (La$_{1-x}$Sr$_{x}$CuO$_{4}$) \cite{imada98}, or in the
disulfides Co$_{1-x}$Fe$_{x}$S$_{2}$ and Ni$_{1-x}$Co$_{x}$S$_{2}$ \cite%
{disulfides}. In the first two examples, the Sr ions create different
potentials in their vicinity which affect the correlated $d$
electrons/holes. In the second set of examples, two different transition
metal ions are located at random positions, creating two different atomic
levels for the correlated $d$ electrons. In both cases the random positions
of different ions break the translational invariance of the lattice, and the
number of $d$ electrons/holes varies. As the composition changes so does the
randomness,  with $x=0$ or $x=1$ corresponding to the pure cases. With
changing composition the system can undergo various phase transitions. For
example, FeS$_{2}$ is a pure band insulator which becomes a disordered metal
when alloyed with CoS$_{2}$, resulting in Co$_{1-x}$Fe$_{x}$S$_{2}$. This
system has a ferromagnetic ground state for a wide range of $x$ with a
maximal Curie temperature $T_{c}$ of $120$ K. On the other hand, when CoS$%
_{2}$ (a metallic ferromagnet) is alloyed with NiS$_{2}$ to make Ni$_{1-x}$Co%
$_{x}$S$_{2}$, the Curie temperature is suppressed and the end compound NiS$%
_{2}$ is a Mott-Hubbard antiferromagnetic insulator with N\'{e}el
temperature $T_{N}=40\;$K.

Our theoretical understanding of systems with strong interactions and
disorder is far from complete. For example, it was realized only recently
that in gapless fermionic systems the soft modes couple to order parameter
fluctuations, leading to different critical behavior in the pure and the
disordered cases \cite{belitz}. A powerful method for theoretical studies of
strongly correlated electron systems is the dynamical mean-field theory
(DMFT) \cite{georges96,pruschke95,vollhardt93}. The DMFT is a comprehensive,
conserving, and thermodynamically consistent approximation scheme which
emerged from the infinite dimensional limit of fermionic lattice models \cite%
{metzner89}. During the last ten years the DMFT has been extensively
employed to study the properties of correlated electronic lattice models.
Recently the combination of DMFT with conventional electron structure theory
in the local density approximation (LDA) has provided a novel computational
tool, LDA+DMFT \cite{anisimov97,held01}, for the realistic investigation of
materials with strongly correlated electrons, e.\,g.\ itinerant ferromagnets %
\cite{lichtenstein01}.

The interplay between local disorder and electronic correlations
can also be investigated within DMFT \cite
{VV+JV92,ulmke95,dobrosavljevic97,laad01,meyer01}. Although
effects due to coherent backscattering cannot be studied in this
way \cite{VV+JV92}, since
the disorder is treated on the level of the coherent potential approximation %
\cite{velicky68}, there are still important physical effects
remaining. In particular, electron localization, and a
disorder-induced metal-insulator transition (MIT), can be caused
by alloy-band splitting.  In the present paper we study the
influence of disorder on the ferromagnetic phase. We will show
that in a correlated system with binary-alloy disorder the Curie
temperature depends non-trivially on the band filling. In the
disordered
one-band Hubbard model we find that for a certain band filling (density) $%
n=N_{e}/N_a$, where $N_{e}$ ($N_a$) is the number of electrons (lattice
sites), disorder can weakly \emph{increase} the Curie temperature provided
the interaction is strong enough. A simple physical argument for this
behavior is presented. We also find that at special band fillings $n\neq 1$
the system can undergo a new type of Mott-Hubbard MIT upon increase of
disorder and/or interaction.

In the following we will study itinerant electron ferromagnetism in
disordered systems, modeled by the Anderson-Hubbard Hamiltonian with on-site
disorder
\begin{equation}
H=\sum_{ij,\sigma }t_{ij}c_{i\sigma }^{\dagger }c_{j\sigma}+ \sum_{i\sigma}
\epsilon_i n_{i\sigma} + U\sum_{i}n_{i\uparrow }n_{i\downarrow },  \label{1}
\end{equation}
where $t_{ij}$ is the hopping matrix element and $U$ is the local  Coulomb
interaction. The disorder is represented by the ionic energies $\epsilon _{i}
$, which are random variables. We consider binary alloy disorder where the
ionic energy is distributed according to the probability density $P(\epsilon
)=x\delta (\epsilon +\Delta /2)+(1-x)\delta (\epsilon -\Delta /2)$. Here $%
\Delta $ is the energy difference between the two ionic energies, providing
a measure of the disorder strength, while $x$ and $1-x$ are the
concentrations of the two  alloy ions. For $\Delta \gg B$, where $B$ is the
band-width, it is known that binary alloy disorder causes a band splitting
in every dimension $d\geq 1$, with the number of states in each alloy
subband equal to $2 xN_a$ and $2(1-x)N_a$, respectively \cite{velicky68}.

We solve (\ref{1}) within DMFT\@. The local nature of the theory implies that
short-range order in position space is missing. However, all dynamical
correlations due to the local interaction are fully taken into account.

In the DMFT scheme the local Green function $G_{\sigma n}$ is given by the
bare density of states (DOS) $N^{0}(\epsilon )$ and the local self-energy $%
\Sigma _{\sigma n}$ as $G_{\sigma n}=\int d\epsilon N^{0}(\epsilon
)/(i\omega _{n}+\mu -\Sigma _{\sigma n}-\epsilon )$. Here the
subscript $n$ refers to the Matsubara frequency $i\omega
_{n}=i(2n+1)\pi /\beta $ for the temperature $T$, with $\beta
=1/k_{B}T$, and $\mu $ is the chemical potential. Within DMFT the
local Green function $G_{\sigma n}$ is determined
self-consistently by
\begin{equation} G_{\sigma n}=-\Bigg\langle
\frac{\int D\left[ c_{\sigma },c_{\sigma }^{\star }\right]
c_{\sigma n}c_{\sigma n}^{\star }e^{{\cal A}_i\{c_{\sigma
},c_{\sigma }^{\star }, {\cal G}_{\sigma }^{-1}\}}}{\int D\left[
c_{\sigma },c_{\sigma }^{\star } \right] e^{{\cal A}_i\{c_{\sigma
},c_{\sigma }^{\star },{\cal G}_{\sigma }^{-1}\}}} \Bigg\rangle
_{\rm dis}, \label{4}
\end{equation}
together with the \textbf{k}-integrated Dyson equation
$\mathcal{G}_{\sigma n}^{-1}=G_{\sigma n}^{-1}+\Sigma _{\sigma
n}$. The single-site action $\mathcal{A}_{i}$ for a site with the
ionic energy $\epsilon _{i}=\pm \Delta /2$ has the form
\begin{eqnarray}
\mathcal{A}_{i}\{c_{\sigma },c_{\sigma }^{\star },\mathcal{G}_{\sigma
}^{-1}\} =\sum_{n,\sigma }c_{\sigma n}^{\star }\mathcal{G}_{\sigma
n}^{-1}c_{\sigma n}-\epsilon _{i}\sum_{\sigma }\int_{0}^{\beta }d\tau
n_{\sigma }(\tau )  \nonumber \\
-\frac{U}{2}\sum_{\sigma }\int_{0}^{\beta }d\tau c_{\sigma }^{\ast }(\tau
)c_{\sigma }(\tau )c_{-\sigma }^{\ast }(\tau )c_{-\sigma }(\tau ),  \label{6}
\end{eqnarray}
where we used a mixed time/frequency convention for Grassmann variables $%
c_{\sigma }$, $c_{\sigma }^{\star }$. Averages over the disorder are
obtained by $\langle \cdots \rangle _{\mathrm{dis}}=\int d\epsilon
P(\epsilon )(\cdots )$.

Since an asymmetric DOS is known to stabilize ferromagnetism in the one-band
Hubbard model for moderate values of $U$ \cite{ulmke98,wahle98,vollhardt00}
we use the DOS of the fcc-lattice in infinite dimensions, $N^{0}(\epsilon
)=\exp [-(1+\sqrt{2}\epsilon )/2]/\sqrt{\pi (1+\sqrt{2}\epsilon )}$ \cite%
{muller-hartmann91}. This DOS has a square root singularity at $\epsilon =-1/%
\sqrt{2}$ and vanishes exponentially for $\epsilon \rightarrow \infty $. In
the following the second moment of the DOS, $W$, is used as the energy scale
and is normalized to unity \cite{comment2}. The one-particle Green function
in Eq.~(\ref{4}) is determined by solving the DMFT equations iteratively\cite%
{ulmke98,wahle98} using Quantum Monte-Carlo (QMC) simulations \cite{hirsh86}.
Curie temperatures are obtained by the divergence of the homogeneous 
magnetic susceptibility \cite{ulmke98,byczuk01}.

\begin{figure}[tbp]
\includegraphics [clip,width=7.cm,angle=-0]{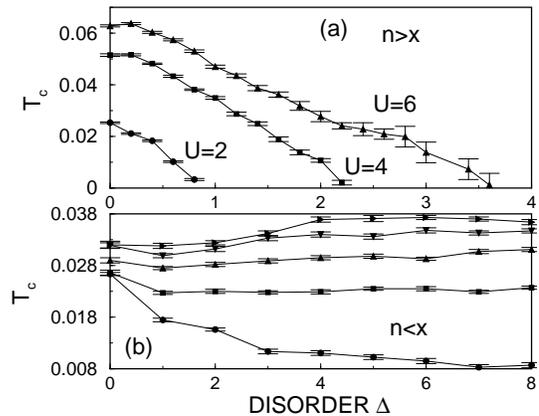}
\caption{Curie temperature $T_c$ as a function of disorder
strength $\Delta$ for band filling $n$ larger [panel (a)] and
smaller [panel (b)] than the ionic concentration $x$ (here
$x=0.5$): (a) $n=0.7$, $U=2,\:4$, and $6$; (b) $n=0.3$,
$U=2,\:3,\:4,\:5$, and $6$ ($U$ increases from bottom to top).
Note the different range of $\Delta$ in both figures. }
\label{fig1}
\end{figure}

We find a striking difference in the dependence of the Curie
temperature $T_c $ on disorder strength $\Delta$ for different
band fillings $n<x$ and $n>x$ (we chose $x=0.5$ for numerical
calculations). At $n=0.7$, the critical temperature $T_c(\Delta)$
decreases with $\Delta$ for all values of $U$ and eventually
vanishes at sufficiently large disorder [Fig.~\ref{fig1}(a)]. By
contrast, at $n=0.3$, $T_c(\Delta)$ weakly decreases with $\Delta$ at small
$U$, but \emph{increases} with $\Delta$ at large values of $U$ 
[Fig.~\ref{fig1}(b)].

\begin{figure}[tbp]
\includegraphics [clip,width=7.cm,angle=-0]{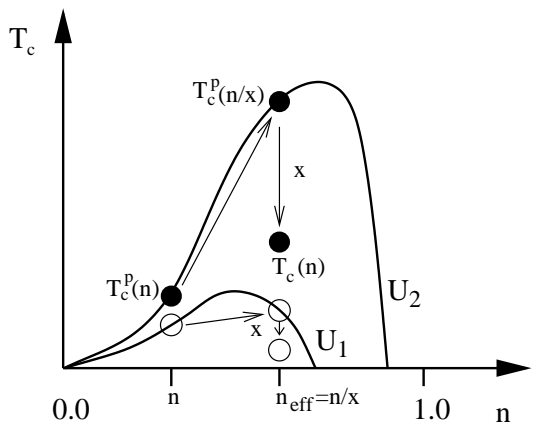}
\caption{Schematic plot explaining the filling dependence of $T_c$
for interacting electrons with strong binary alloy disorder.
Curves represent $T_c^{\mathrm{p}}$, the Curie temperature for the
pure system, as
a function of filling $n$ at two different interactions $U_1\ll U_2$ 
(cf.~\protect\cite{ulmke98}). For $n\lesssim x$, $T_c$ of the
disordered system can be obtained by transforming the open (for
$U_1$) and the filled (for $U_2$) point from $n$ to
$n_{\mathrm{eff}}$, and then multiplying $T_c^{\mathrm{p}}(n/x)$
by $x$ as indicated by arrows. One finds
$T_c(n)<T_c^{\mathrm{p}}(n)$ for $U_1$, but
$T_c(n)>T_c^{\mathrm{p}}(n)$ for
$U_2$. This difference originates from the non-monotonic dependence of $T_c^{%
\mathrm{p}}$ on $n$.}
\label{fig2}
\end{figure}

As will be explained below, this striking difference originates
from three distinct features of interacting electrons in the
presence of binary alloy disorder:

i) $T_c^{\mathrm{p}}\equiv T_c(\Delta=0)$, the Curie temperature
in the pure case, depends non-monotonically on band filling $n$. 
Namely, $T_c^{\mathrm{p}}(n)$ has a maximum at some filling
$n=n^{\ast}(U)$, which increases as $U$ is increased
\cite{ulmke98}; see Fig.~\ref{fig2}.

ii) In the alloy disordered system the band is split
\cite{velicky68} when $\Delta\gg W$. As a consequence, for $n<2x$
and $T\ll \Delta$ electrons only occupy the lower alloy subband
while the upper subband is empty. Effectively, one can therefore
describe this system by a Hubbard model  mapped onto the lower
alloy subband. Hence, it corresponds to a \emph{single} band with the \emph{%
effective} filling $n_{\mathrm{eff}}=n/x$. It is then possible to
determine
$T_c$ from the phase diagram of the Hubbard model without disorder \cite%
{ulmke98}.

iii) The disorder leads to a reduction of $T_c^{\mathrm{p}}(n_{\mathrm{eff%
}})$ by a factor $x$, i.\,e.\ we find
\begin{equation}
T_c(n)\approx xT_c^{\mathrm{p}}(n/x)  \label{equation}
\end{equation}
when $\Delta \gg W$  \cite{comment3}. Hence, as illustrated in
Fig.~\ref{fig2}, $T_c$ can be
determined by $T_c^{\mathrm{p}}(n_{%
\mathrm{eff}})$. Surprisingly, then, it follows that, if $U$ is
sufficiently strong, the Curie temperature of a disordered system
can be higher than that of the corresponding pure system 
[cf.~Fig.~\ref{fig2}]!

\begin{figure}[tbp]
\includegraphics [clip,width=7.cm,angle=-0]{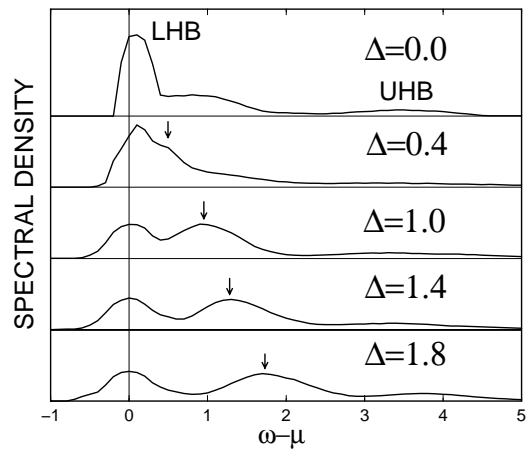}
\caption{Spectral density for different disorder strengths $\Delta$ at $n=0.3
$ and $U=4$ as obtained by the maximal entropy method from QMC data at $%
T=0.071$. The position of the lower/upper Hubbard subbands
(LHB/UHB) are almost unaffected by the disorder, while the upper
alloy subband shifts to the right as indicated by arrows. }
\label{fig3}
\end{figure}

To illustrate the alloy band splitting in the presence of strong
interactions discussed above [see (ii)] we calculate the
spectral density from the QMC results by the maximal entropy method \cite%
{sandvik98}. The results in Fig.~\ref{fig3} show the evolution of the
spectral density in the paramagnetic phase at $U=4$ and $n=0.3$. At $\Delta=0
$ the lower and upper Hubbard subbands can be clearly identified. The
quasiparticle resonance is merged with the lower  Hubbard subband due to the
low filling of the band, and is reduced by the finite temperature. At $%
\Delta>0$ the lower and upper alloy subbands begin to split off. A similar
behavior was found at $n=0.7$. The separation of the alloy subbands in the
correlated electron system for increasing $\Delta$ is one of the
preconditions [cf.~(ii)] for the enhancement of $T_c$ by disorder when $n<x$%
, as discussed above.

The splitting of the alloy subbands and, as a result, the changing of the
band filling in the effective Hubbard model implies that $T_c$ vanishes for $%
n>x$. Namely, in the ferromagnetic ground state each of the alloy subbands
can accommodate only $xN_a$ and $(1-x)N_a$ electrons, respectively.
Therefore, if the ground state of the system were ferromagnetic the upper
alloy subband would be partially occupied for all $n>x$. This would,
however, increase the energy of the system by $\Delta$ per particle in the
upper alloy subband. Therefore,  in the $\Delta \gg U$ limit the
paramagnetic ground state is energetically favorable. This explains why $T_c$
vanishes at $n=0.7$, as found in our QMC simulations [Fig.~\ref{fig1}(a)].
Our conclusion that $T_c$ vanishes for $n_{\mathrm{eff}}=n/x>1$ when $%
\Delta\gg W$ is consistent with the observation in \cite{ulmke98} that 
there is no ferromagnetism for $n>1$ in the Hubbard model without disorder
on fcc-lattice in infinite dimensions.

\begin{figure}[tbp]
\includegraphics [clip,width=7.cm,angle=0]{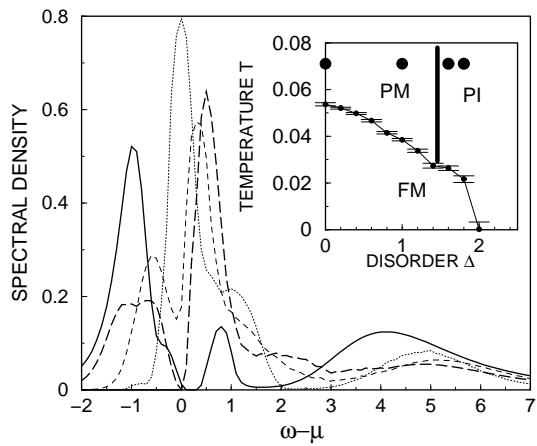}
\caption{Spectral density for disorder strengths $\Delta=0,\; 1,\;1.6$, and $%
1.8$ (dotted, dashed, long-dashed and solid curves, respectively) at $n=0.5$
and $U=6$ as obtained by the maximal entropy method from QMC data at $T=0.071
$. For $\Delta\gtrsim 1.6$ a Mott-Hubbard gap-like structure develops around
the Fermi level. Inset: $\Delta-T$ phase diagram of the binary alloy Hubbard
model on the fcc-lattice in infinite dimensions at $U=6$; PM - paramagnetic
metal, PI - paramagnetic insulator-like phase, FM - ferromagnetic metal.
Points with error bars represent the Curie temperatures obtained from QMC
simulations; the solid line is a guide for the eye only. The thick line
indicates the phase boundary between the PM and PI phases (see text).
Circles: parameter values ($\Delta,T$) corresponding to the spectral
densities shown in the main panel. }
\label{fig4}
\end{figure}

The filling $n=x$ is very particular because a new MIT of the Mott-Hubbard
type occurs. Namely, when $\Delta$ increases (at $U=0$), the non-interacting
band splits,  leaving $2 xN_a$ states in the lower and $2(1-x)N_a$ states in
the upper alloy subband. Effectively, it means that  at $n=x$ the lower
alloy subband is half filled ($n_{\mathrm{eff}}=1$). Consequently, a
Mott-Hubbard MIT occurs in the lower alloy subband at sufficiently large
interaction $U$ \cite{comment5}. In fact, for $\Delta\gg U$ we may infer a
critical value $U_c=1.47 W^*$ at $T=0$ from the results of Refs.~\cite%
{moeller95,bulla99}, where $W^*$ is the renormalized bandwidth of the lower
alloy subband. 
Furthermore, from the analogy of this MIT with that in the pure case \cite%
{bulla01} we can expect a discontinuous transition  for $T\lesssim
T^*\approx 0.02 W^*$, and a smooth crossover for $T\gtrsim T^*$. From the
results shown in Fig.~\ref{fig4} it follows that $T^*<0.071$, since for $%
T=0.071$ and $U=6$ a gap-like structure develops in the spectrum at $%
\Delta\approx 1.6$, implying a smooth but rapid crossover from a metallic to
an insulator-like phase \cite{comment6}.

The MIT described above is not obscured by the onset of antiferromagnetic
long-range order because in infinite dimensions the fcc-lattice is
completely frustrated \cite{muller-hartmann91}. Hence the insulator is
paramagnetic. The transition therefore occurs between a paramagnetic
insulator (PI) at high $T$ and a \emph{ferromagnetic metal} (FM) at low $T$,
at least at large $U,$ as shown in the inset of Fig.~\ref{fig4}. The
actual boundary between the paramagnetic metal (PM) and the paramagnetic
insulator-like phase has not yet been determined. The thick line in the
inset of Fig.~\ref{fig4} indicates the approximate position  of the phase
boundary between the PM and PI phases. We note that at the point where this
boundary meets the FM phase the slope of the transition line $T_c(\Delta)$
changes.

In summary, we showed within DMFT that the interplay between binary-alloy
disorder and electronic correlation can result in unexpected effects, such
as the enhancement of the transition temperature $T_c$ for itinerant
ferromagnetism by disorder, and the occurrence of a Mott-Hubbard type MIT
off half-filling. An observation of these effects requires good control of
the system parameters over a wide range as was recently shown to be possible
in experiments with optical lattices \cite{greiner02}.

We thank B.\ Velicky for valuable correspondence. KB is grateful to
R.\ Bulla and K.\ Wysoki\'nski for discussions, and to G.\ Keller for
computer assistance. This work was supported by a Fellowship of
the Alexander von Humboldt-Foundation (KB), and through SFB 484 of
the Deutsche Forschungsgemeinschaft.




\begin{references}
\bibitem{comment2}Exponential tails in the DOS prohibit the formation of a
genuine disorder and/or correlation induced gap in the spectrum.
However, a ``soft gap'', where the spectrum vanishes at a single point, is
permitted.
\bibitem{comment3}
Eq.\ (4) can be derived within Hartree-Fock theory. We find it to
be valid even at strong interactions.
\bibitem{comment5} The same  holds true when $n=1+x$, i.\,e.\ when the upper
  alloy subband is half-filled.
\end{references}

\end{document}